\documentclass[prl,
twocolumn,
aps,psfig]{revtex4} %/MH

\usepackage{graphicx,color,epsfig,rotate}

\begin{document}
\draft

\title{
High-Field Pauli-limiting behavior and Strongly 
Enhanced Upper Critical Magnetic Fields
 near the Transition Temperature of an Arsenic-Deficient 
 LaO$_{0.9}$F$_{0.1}$FeAs$_{1-\delta}$
 Superconductor}

\author{ G.~Fuchs, S.-L.~Drechsler$^{*}$, N.~Kozlova, G.~Behr, A.~K\"ohler, 
J.~Werner, K.~Nenkov, 
R.~Klingeler, J.~Hamann-Borrero, C.~Hess,
A.~Kondrat, 
M.~Grobosch, A.~Narduzzo, M.~Knupfer,
 J.~Freudenberger, B.~B\"uchner, 
and L.~Schultz }

\affiliation{IFW Dresden, P.O.~Box 270116, D-01171 Dresden,
Germany}

\date{\today}
\begin{abstract}
We report upper critical field $B_{c2}(T)$ data for disordered (arsenic-deficient)
LaO$_{0.9}$F$_{0.1}$FeAs$_{1-\delta}$ in a wide temperature and 
magnetic
field range up to 
47~T.
 Because of the large
 linear 
% initial 
slope of $B_{c2}\approx$~-5.4 to -6.6T/K 
near $T_c\approx$ 28.5~K the $T$-dependence of the in-plane 
$B_{c2}(T)$ shows a  
%clear 
flattening 
%already 
near 23~K above 30~T 
which 
%is intepreted as 
points to Pauli-limited behavior with $B_{c2}(0)\approx$~63 to 68~T. 
Our results are discussed in terms of disorder effects within
conventional
 and unconventional 
superconducting pairings.  

\end{abstract}


\maketitle

The recent discovery of 
relatively 
high transition
temperatures $T_c$ 
in LaO$_{1-x}$F$_{x}$FeAs \cite{kamihara08} 
has established a new
family of superconductors. Since
the usual $el$-$ph$ mechanism has been 
ruled out by a much too
weak coupling strength $\lambda \leq$~0.2
 \cite{boeri}, 
a variety of nonstandard 
mechanisms mostly
 involving spin fluctuations 
has been proposed \cite{xu,mazin,korshunov}.
 Naturally, 
 our knowledge about these fascinating systems
is still poor at present. 
Together with the controversially discussed 
unconventional ($p$- or $d$-wave) vs conventional (extended $s$-wave) 
symmetry and magnitude of the superconducting order parameter $\Delta$
(gap) \cite{chengap,dubrowka}, 
the upper critical field $B_{c2}(T)$ and its slope
near $T_c$ are  fundamental quantities
characterizing the superconducting state. Due to the involved
Fermi velocities $v_f$ in the clean limit it provides insight into 
the underlying 
%yet unknown 
electronic 
structure which might be affected by a complex interplay of
correlation effects \cite{haule,haule2}   
and 
the vicinity
 of competing magnetism 
\cite{dong,weng,singh}.
Controlled disorder provides insight into relevant scattering 
processes and in the symmetry of the pairing
 since any
unconventional pairing in the 
sense of $T_c$ and $dB_{c2}/dT |_{T_c}$ 
is expected to be suppressed by 
strong disorder 
 \cite{lin,petrovic,mackenzie,radtke}:
\begin{equation}
-\ln \left( \frac{T_c}{T_{c0}} \right)=\psi \left( \frac{1}{2}+
\frac{\beta T_{c0}}{2\pi T_c} \right) -\psi \left( \frac{1}{2}\right), 
\end{equation}
where $\psi (x)$ is the digamma function and $\beta$ is the 
 strong-coupling 
pair-breaking parameter
$\beta = \Omega^2_p \rho_0/8\pi (1+\lambda) T_{c0}$ 
which is 
related to the residual resistivity $\rho_0$ 
and the plasma energy
$\Omega_p$ in the $(a,b)$-plane. Some, 
but much weaker, suppression might occur
also in the anisotropic or multiband $s$-wave case
since the scattering may smear out the gap 
anisotropy. However, it will be shown that surprisingly 
nothing similar happens in our case.

For low applied fields
 rather different slopes  
$dB_{c2}/dT\approx$~-1.6 T/K up to -2 T/K at $T_c\approx$~26~K
 \cite{sefat,hunte} and up to -4~T/K at $T_c\approx$ 20~K~ \cite{chen1}
have been reported
for the As stoichiometric La based compounds. Here, 
we report with $dB_{c2}/dT\approx$~-5.4 to -6.6 T/K, 
to our knowledge
 one of the highest 
slopes of $B_{c2}$ near $T_c$
 observed so far
%both 
for the La-series. Another interesting issue of high-field studies
considered here is the possibility of Pauli-limiting (PL) behavior.
Triplet $p$-wave pairing 
%\cite{lee,dai} 
or  strong coupling ($B_{c2}(0)\leq$ 50~T) 
would naturally explain the reported absence of PL \cite{hunte}. 
In this context
it is important that we succeeded to detect PL behaviour for our specific
sample. It
 points to $B_{c2}(0)$ values
%$ \approx$ 63~\textcolor{red}{to 68}~T 
being much below often-used 
%naive 
WHH (Werthamer-Helfand-Hohenberg) \cite{WHH} 
%($\sim$ 106~T in our case) 
based estimates
.
After presenting 
%structure, resistivity and  
various data 
 which deviate 
from those of Ref.~\onlinecite{hunte} as well as 
from 
%all
our non-deficient As 
samples 
%studied 
%at the IFW Dresden 
%so far 
\cite{luetkens,klauss,drechsler,graefe,hess}, we will discuss our
 $B_{c2}(T)$
%high-field
results in 
the light of these 
more general issues.

A 
polycrystalline sample of LaO$_{0.9}$F$_{0.1}$FeAs$_{1-\delta}$ was prepared
as described in our 
previous work \cite
{luetkens,klauss,drechsler,graefe,hess} and e.g.\ in 
Ref.~
\onlinecite{zhu}. However, in contrast to that work here a Ta 
foil was used to wrap the pellets. Ta acts as an As getter at 
high $T$ forming a solid solution of about 9.5~at\% As in Ta 
with a small layer of Ta$_2$As and TaAs on top of the foil. 
This leads 
to an As loss in the pellets.
 The annealed pellets were 
ground and polished and the
local composition of the resulting samples was investigated by
wavelength-dispersive x-ray spectroscopy (WDX) in a scanning
electron microscope. 
 The 
sample consists of 1 to~20~$\mu$m sized grains of 
LaO$_{0.9}$F$_{0.1}$FeAs$_{1-\delta}$ ($\delta \sim $ 0.05 to 0.1)  
where the F content slightly fluctuates between different grains. 
A powder x-ray diffraction study 
with a Rietveld refinement of the main phase
yields slightly enhanced lattice constants: $a$=~4.02819~\AA \ and
$c$=~8.72397~\AA \ compared to 
$a$=~4.02043(4.02451)~\AA \ and
$c$=~8.69552(8.70995)~\AA \ for
our cleanest sample with $T_c=$~26.8~K and an ordinary sample
with $T_c=26$~K, respectively \cite{hess}.
 The latter are
 almost
As-stoichiometric samples with the same
F$_{0.1}$-content.
 Surprisingly, the lattice constants of the ADS 
(As deficient sample)
 are close 
to those for underdoped samples near the border of magnetism and 
superconductivity in stoichiometric samples. 
The reduced charge of the anionic As layers suggests
less attraction between them and the adjacent "cationic" central Fe and the 
charge reservoir LaO$_{0.9}$F$_{0.1}$ 
layers. This might explain the increase of $c$ and due to the weakened
Madelung potential also a reduced doping of the Fe-layer.

The electrical resistance was measured for a 
\begin{figure}[t]
\includegraphics[width=7cm]{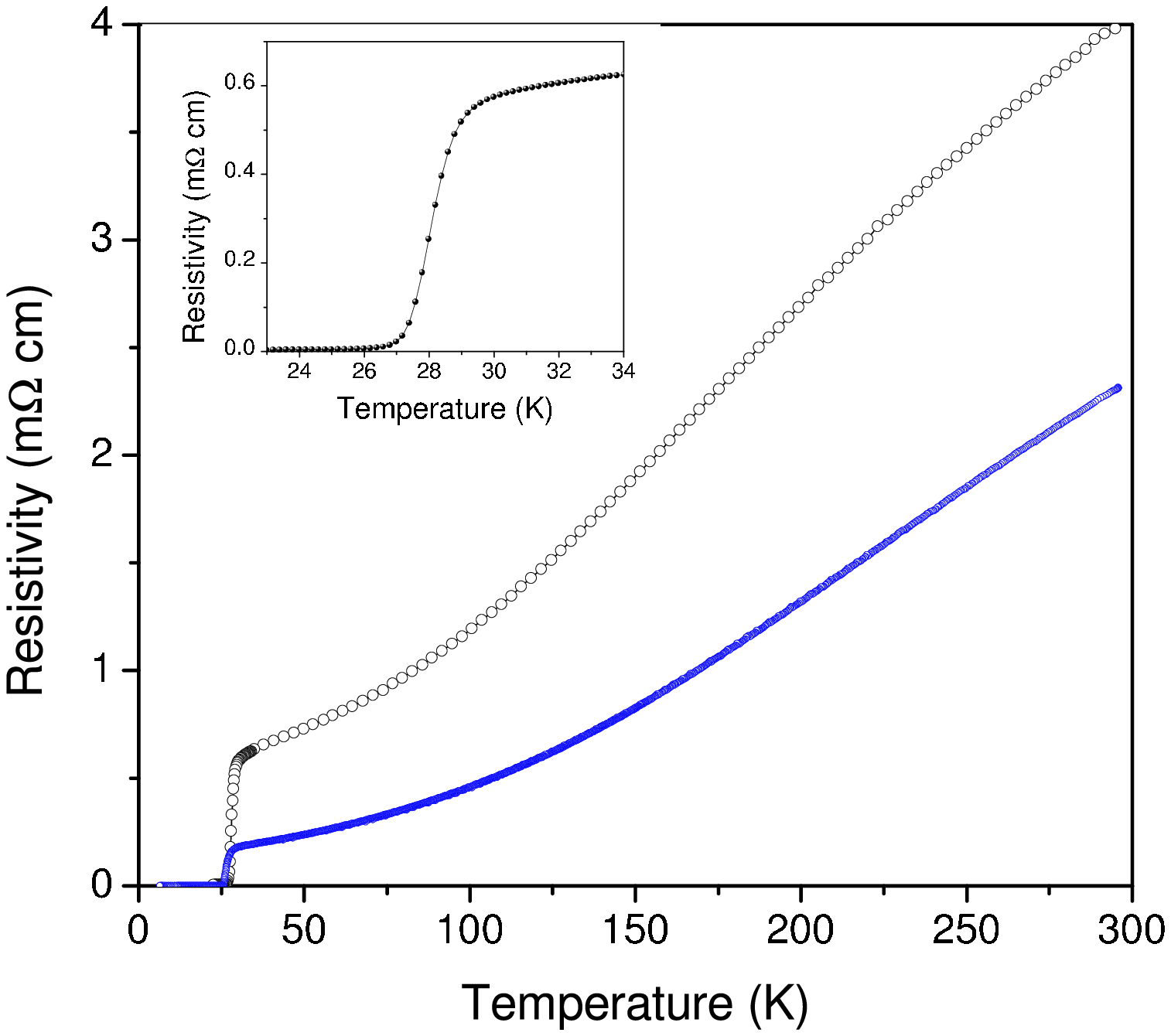}
\caption{(Color online) Resistivity
at zero magnetic field of the 
LaO$_{0.9}$F$_{0.1}$FeAs$_{1-\delta}$, $\delta\approx 0.05$ to 0.1
 ADS studied in the present paper. 
The inset shows 
the 
%low-$T$ 
resistivity near $T_c$.
Solid line: a cleaner almost stoichiometric sample \cite{hess}.
} \label{f1}
\end{figure}
\begin{figure}[b]
\includegraphics[width=7.5cm]{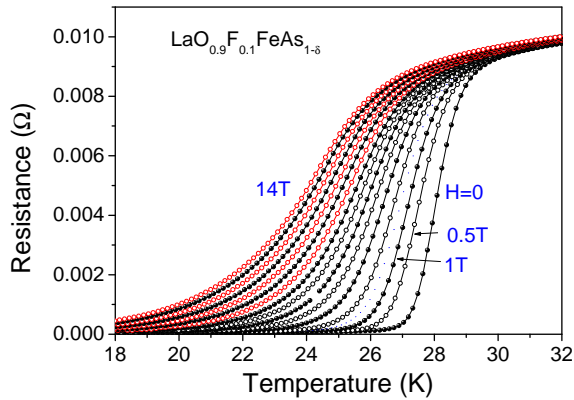}
\caption{(Color online) $T$-dependence of the resistance $R$ for the ADS
for various dc fields up to 14~T. Between 1 and 14~T, the applied magnetic 
field was increased in steps of 1~T.}
\label{f2}
\end{figure}
plate-like LaO$_{0.9}$F$_{0.1}$FeAs$_{1-\delta}$ 
ADS with nominal dimensions 
$3 \times 2.6 \times 0.53$~mm$^3$    
using the standard four-point method. 
Its resistivity $\rho (T)$ is shown in Fig.~1. The 
resistivity of this ADS in the normal state at 30~K, with 
about 0.6~m$\Omega$cm, clearly exceeds that reported for other 
LaO$_{0.9}$F$_{0.1}$FeAs samples. 
Nevertheless, our 
ADS was found to exhibit a rather high $T_c$-value of 28.5~K 
defined at 90\% of $\rho$ in the normal state 
and a relatively sharp transition width (see inset of Fig.~1) which 
excludes an anomalous inhomogeneity. 
The low-$T$ region above $T_c$ with $\rho \propto T^2$,
ascribed to a pronounced $el$-$el$ scattering  
was regarded as evidence for a 
standard Fermi liquid picture \cite{sefat}, has been 
somewhat narrowed from 225~K to $T \leq$ 175~K for this ADS \cite{remrho}.
The $\rho(T)$-dependence of this ADS resembles that of underdoped
stoichiometric samples in the range of 0.05 to 0.07 
F-content \cite{hess}. 
Since each As site is surrounded by four 
Fe sites, the effect of 
even few As vacancies might be drastic. Thus, a substantial
 shortening of the mean free path $l$ up to few lattice constants $a$ 
as estimated
below from the observed enhanced field slope at $T_c$ seems to be quite 
reasonable. 

\begin{figure}[t]
\includegraphics[width=7.5cm]{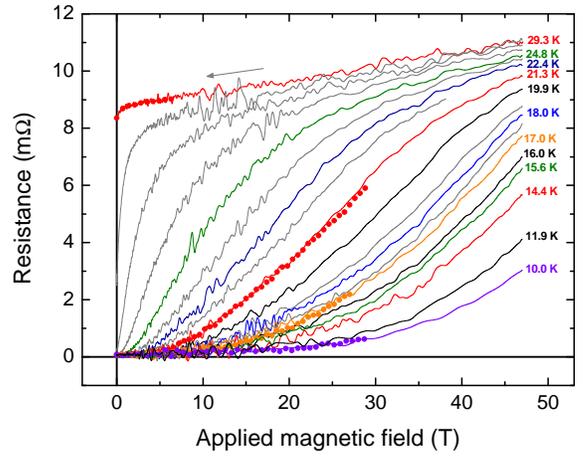}
\caption{(Color online) Field dependence of the resistance at 
fixed $T$ (see legend) measured in pulsed fields.  Lines: 
measurements up to 47~T; symbols: measurements shown for four selected 
$T$-values.}
\label{f3}
\end{figure}

In Fig.~2, the electrical resistance of the 
%investigated 
studied ADS is plotted 
$vs$.~$T$ for applied dc fields up to 14~T. Resistance data 
{\it vs.}~applied 
field  of this ADS measured in pulsed fields up to 50~T are 
plotted in 
Fig.~3. Gold contacts (100 nm thick) were made by sputtering
in order to provide a low
contact resistivity and therefore to avoid possible heating effects
in the high-field measurements performed in the 50~T magnet of the 
Dresden-High-Magnetic-Field-Laboratory \cite{lab}. 
The agreement between measurements up to 29 T and 47 T 
confirms that our data are not affected by sample heating.
At high 
fields  a substantial broadening of 
the transition curves is observed as shown in both figures. It
stems from the large anisotropy of $B_{c2}(T)$ 
expected
 for 
layered compounds as here \cite{mazin,singh}.
$B_{c2}$ was determined as in Ref.~\onlinecite{hunte} from the 
onset of 
superconductivity defining it at  90\% of $R_N$, the resistance in 
the normal 
state. Within a second approach
the magnetoresistance in the normal state was taken into account 
and $R_N(T)$ has been described as explained in Ref.\ \onlinecite{remrho}.
For this modified definition of $B_{c2}$ one gets somewhat higher
$B_{c2}$-values and also higher slopes near $T_c$.

Generally, the $B_{c2}$ values
 from $0.9R_N$ data refer to those grains which 
are oriented 
with their $ab$-planes along the applied field. The $B_{c2}^{ab}(T)$
curve
 of our ADS is shown in 
Fig.~4. The comparison of the data from dc and 
pulsed field measurements in the 
field range up 
to 14~T confirms that both 
data sets do well agree. The $B_{c2}(T)$-curve 
in Fig.~4 shows a surprisingly 
steep 
$dB_{c2}/dT|_{T_c}$~=~-5.4 T/K (-6.6 T/K within the second 
approach described above) which exceeds the slopes reported for cleaner
 non-ADS 
samples by more than a factor of two 
\cite{zhu,hunte,sefat}. This 
%might 
points to strong impurity scattering in our ADS 
in accord with its enhanced 
resistivity at 30~K. For applied fields up to 
about 30~T, 
the $B_{c2}(T)$ data can be well described by the standard WHH model \cite{WHH}. 
Using $dB_{c2}/dT$~=~-5.4 T/K (-6.6 T/K) 
and $T_c$~=~28.5 K (28.8 K), 
this model predicts 
$B_{c2}^*(0)$ = -0.69~$T_c (dB_{c2}/dT)_{T_c}$~=~106~T 
(131 T) at $T$~=~0. 
However, for applied fields above 30~T or 
below 23~K increasing 
deviations of the $B_{c2}(T)$
 data 
from the WHH-curve are clearly visible. 
They increase with applied field.
Notice that the resulting difference between the measured
$B_{c2}(T)$ and $B_{c2}^*$ is comparable for both definitions of 
the upper critical field (Fig.\ 5).
The flattening of $B_{c2}(T)$ 
at high 
fields points to its 
limitation by 
the Pauli spin paramagnetism. 
This effect 
is measured 
in the WHH model 
by the Maki parameter $\alpha$
\begin{equation}
\alpha=\sqrt{2}B^*_{c2}(0)/B_p(0),
\label{maki}
\end{equation}
where $B_p(0)$ is the paramagnetically limited field \cite{orlando}:
\begin{equation}
B_p(0)\mbox{[Tesla]}=1.86\eta_{\mbox{\tiny eff}}(\lambda)T_c\mbox{[K]},
\end{equation}
where $\eta_{\mbox{\tiny eff}}=
(1+\lambda)^\varepsilon\eta_{\Delta}\eta_{B_{c2}}(1-I)$ is a 
correction to BCS due to the $e$-$boson$ and $e$-$e$ 
interaction ($I=N(0)V$ is the Stoner factor)
 \cite{orlando,schossmann}. 
$\eta_{\mbox{\tiny eff}}$ depends on the $el$-$boson$ coupling 
constant $\lambda$ and $\varepsilon$~=~0.5, 1 according to 
Refs.~\onlinecite{orlando,schossmann}, respectively.
Due to PL, $B_{c2}(0)$
 is lowered: 
%to
\begin{equation}
B^p_{c2}(0)=B^*_{c2}(0)/\sqrt{1+\alpha^2}.
\end{equation} 
Ignoring  weak 
spin-orbit scattering for the sake of simplicity, 
we get  $\alpha$~=~1.31 
and $B_{c2}^p(0)$~=~63~T (68 T) from Eqs.~(2)-(4), 
using $\lambda$~=~0.5 \cite{drechsler} 
for a representative value of the $el$-$boson$ coupling 
constant for 
LaO$_{0.9}$F$_{0.1}$FeAs$_{1-\delta}$,
$\eta_{eff}$~=~2.09, and $B_{c2}^*(0)$~=~106~T (131 T)
 without PL. 
The $B_{c2}^p(T)$ line plotted in Fig.~4 is based on Eq.~(4) 
and was obtained by replacing $B_{c2}^*(0)$ entering its 
numerator and $\alpha$ by $B_{c2}^*(T,\alpha=0)$ of the WHH model.  
This rough approximation of $B_{c2}^p(T)$  
has been used to illustrate the PL in the studied ADS.  
\begin{figure}[t]
\includegraphics[width=8.5cm]{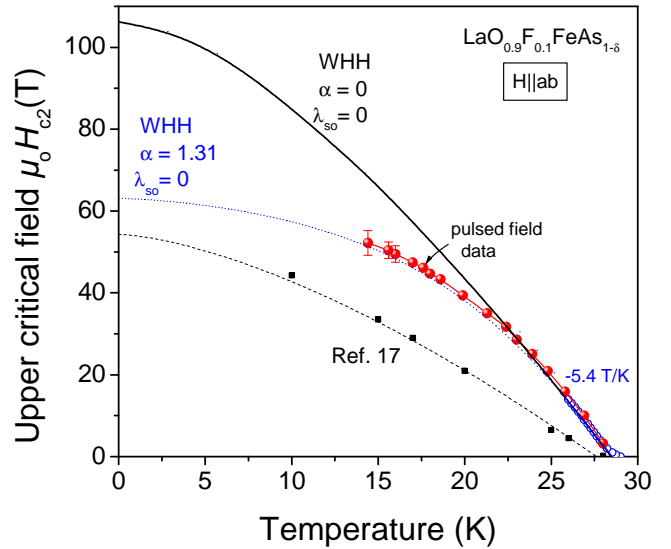}
\caption{
$B^{ab}_{c2}$ {\it vs.}~$T$. Data 
from dc 
($\circ $
%open circles
) and pulsed field 
measurements (
%filled circles
$\bullet $). 
The last three "data points" above 47~T  
are obtained from Fig.~3 by 
a linear extrapolation of 
$R(B)$ $<$ 0.9 $R_N$ to $R(B)$ = 0.9$R_N$. 
Dashed line: 
data from Ref.~\onlinecite{hunte}; 
solid line: WHH-model 
without
 PL. Dotted line: $B_{c2}^*(T)$ for $\alpha$~=~1.31 
without spin-orbit scattering.} 
\label{f4}
\end{figure}

High $B_{c2}(0)$-values and similarly
 for $dB_{c2}/dT|_{T_c}$ can be achieved 
by: (i) strong coupling, (ii) small Fermi velocities
 $v_f$, and (iii) strong 
intra-band scattering. Strong coupling can be excluded empirically 
for related systems \cite{drechsler,chengap}. 
Note that unconventional
 triplet $p$- or $d$-wave pairing 
\cite{petrovic,lin,mackenzie} 
and (iii) can not be reconciled.
Hence, singlet extended 
$s$-wave pairing remains as a reasonable scenario. 
As mentioned above the residual resistivity $\rho_0$ of our ADS 
is enhanced 
compared to that of  clean samples (Fig.~1) examined so far in our studies. 
This points to strong intraband scattering.
 Adopting in accord with the single 
isotropic gap found for SmFeAsO$_{0.85}$F$_{0.15}$ \cite{chengap}
an effective single-band $s$-wave picture \cite{shulga}, $B^c_{c2}(0)$ 
(and similarly for $B^{ab}_{c2}(0)$), written
 in convenient units, ignoring PL
%effects for the sake of simplicity 
reads
 in the clean (cl) and general case:
\begin{equation}
B^c_{c2,cl}(0)\mbox{[T]}=0.0237 \left( 1+\lambda \right) ^{2.2}T^2_c\mbox{[K]}/
v^2_f\mbox{[10$^5$m/s]},
\end{equation}
\begin{equation}
B^c_{c2}(0)=B_{c2,cl}(0)\left[1+0.13\gamma_{imp}\mbox{[K]}/(T_c (1+\lambda )) 
\right],
\end{equation}
where $\lambda$ is the relevant $e$-$boson$  coupling 
constant, $\gamma_{imp}$ is the scattering rate, and $v_f$ denotes the in-plane
averaged Fermi velocity. 
The mass anisotropy $\gamma^2=M/m$ which affects $B^{ab}_{c2}$ varies 
according 
to the LDA-predictions
\cite{drechsler,mazin,singh} 
between 6.2 and 15. 
From the 5\% level of the resistivity
 transition curves 
(see Fig.~5) 
\begin{figure}[b]
\includegraphics[width=8.0cm]{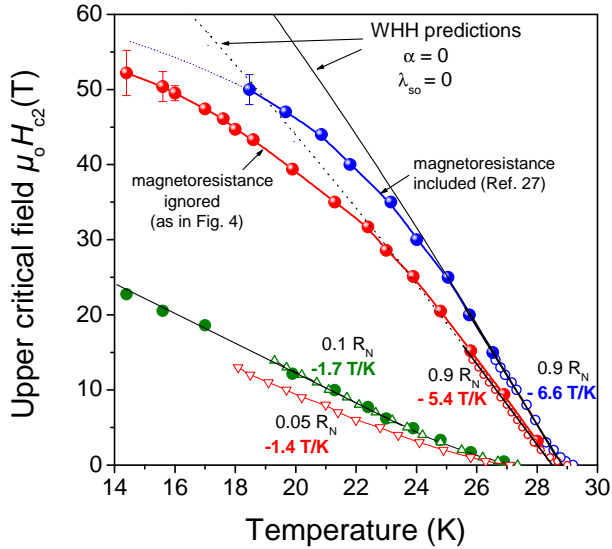}
\caption{
(Color online) Upper critical field {\it vs}. $T$ from 90\%, 
10\% and 5\% of the normal state resistance $R_N$. Open (filled) symbols: dc 
field data - Fig.~2 (pulsed field data - Fig.~3).}
\label{f5}
\end{figure}
a lower limit of about 3.7 for $\gamma=B^{ab}_{c2}/B^c_{c2}$ 
is estimated close to  
$\gamma \approx$ 4.5 for the first single 
crystal of the whole class ( NdO$_{0.82}$F$_{0.12}$FeAS 
\cite{jia} 
). 
The reduced $\gamma$ of these samples compared with 
Refs.~\onlinecite{dubrowka,weyeneth} might be ascribed also to disorder
provided in both cases a less anisotropic Fermi surface sheet is involved like
in MgB$_2$ \cite{gurevich}. 
The 
$B_{c2}(T)$~data for the 10\% $R_N$
 criterion show no PL up to 33~T. 
Hence, 
for $B^c_{c2}$ 
no PL is expected.
 To simulate the substantially
enhanced $dB_{c2}/dT$ at $T_c$ which, estimated in the weak-coupling 
regime, is
 $\propto B_{c2}(0)$
 \cite{bulayevskii}, we adopt using 
Eq.~(6) $\gamma_{imp}\approx 50$~meV, i.e.~a situation intermediate
between clean and dirty limit. The ratio between 
the BCS-coherence length $\xi_0$ and the mean free path $l$
would be 3.6 justifying roughly the above use of 
the dirty limit theory valid 
for $\xi_0/l \gg 1$.  The increase of $T_c$ is difficult to understand
within a simple $s$-wave scenario.
For instance, in V$_3$Si, Nb$_3$Sn
 and MgB$_2$ the slopes 
of $B_{c2}$ raise with increasing disorder
measured by $\rho_0$, whereas their  $T_c$'s slightly 
decrease \cite{orlando,gurevich}.
Here the raise of $T_c$ might result
from (i) an enhanced density of states $N(0)$ due-to disorder,
if the Fermi energy is
 located
in a tail of a broadened peak of $N(E)$, (ii) spin fluctuation $s$-wave 
pairing supported by weak $el$-$ph$ interaction being pair-breaking
in $p$ and $d$ channels,
 and/or (iii) suppression of 
 competing (local) antiferromagnetism by a reduction of the nesting. 
If the pairing in the clean samples is indeed
 unconventional, 
"nonmonotonous" disorder dependencies should first 
cause a weakening of that 
pairing
 followed by an improved $s$-wave pairing as probably observed
 here. 

To summarize, we reported a high-field study of a 
LaO$_{0.9}$F$_{0.1}$FeAs$_{1-\delta}$ sample with
 improved 
superconductivity. It exhibits a rather high slope 
$dB_{c2}/dT|_{T_c}\approx$~-5.4 T/K 
to -6.6 T/K
% in 
%dependence
depending
 on the used definition of $B_{c2}(T)$. In all cases at
lower $T$ a flattening of 
the $B_{c2}(T)$-curve points
to 
%Pauli limiting 
PL behavior with $B_{c2}(0)\approx$ 63~T to 68 T
 extrapolated.
We ascribe this behavior to disorder
effects in an extended $s$-wave state.
Thus, controlled disorder provides a useful tool to study
the pairing symmetry of
these novel 
%layered 
superconductors. 
%It would be also helpful to clarify the pairing symmetry. 
%these samples.
In view of the achieved improved 
$dB_{c2}/dT |_{T_c}$ 
at a higher $T_c$-value,
the introduction of As vacancies or of other defects opens new routes for 
optimizing their properties. Its study 
should provide also a deeper insight into the specific role
of As 4$p$ orbitals played in the formation of quasiparticles relevant 
for the physical  
%superconducting and normal state 
properties 
%\cite{marsiglio} 
%as well as into that of 
including
the magnetic excitations.
The PL 
%Pauli limiting 
found here suggests to continue 
%high-field 
measurements at least up to 70~T in order to elucidate, 
whether there is still much room for increasing $B_{c2}$ beyond that range.
Apparently, the solution of this  problem 
will affect
%is of great importance for 
the evaluation of
 future high-field applications.

We thank M.~Deutschmann, R.~M\"uller, S.~Pichl, 
R.~Sch\"onfelder, and S.~M\"uller-Litvanyi  for 
technical assistance
. Discussions with  H.~Eschrig, 
A.~Gurevich, I.~Ma-
zin,
and  K.~Koepernik 
are grateful acknowledged.


\begin{thebibliography}{00}
\bibitem[*]{dre}Corresponding author.\\ 
drechsler@ifw-dresden.de

\bibitem{kamihara08}Y.~Kamihara {\it et al.}, 
J.\ Am.\ Chem.\ Soc.\ {\bf 130}, 3296 (2008).
\bibitem{boeri}L. Boeri {\it et al.}, 
Phys.\ Rev.\ Lett.\ {\bf 101 }, 026403 (2008). 
\bibitem{xu}G. Xu {\it et al.}, Europhys.\ Lett.\ {\bf 82} 67002 (2008).
\bibitem{mazin}I.I. Mazin {\it et al.}, Phys.\ Rev.\ Lett.\ 
{\bf 101}, 057003 (2008).
\bibitem{korshunov}M.M.~Korshunov and I.~Eremin, arXiv: 0804.1793.

\bibitem{chengap}T.Y.~Chen {\it et al.}, Nature {\bf 453}, 
1224 (2008). 
\bibitem{dubrowka}A.~Dubroka {\it et al.}, 
Phys.\ Rev.\ Lett. {\bf 101}, 097011 (2008).
\bibitem{haule}K.~Haule {\it et al.}, Phys.\ Rev.\ Lett.\ {\bf 100}, 226402 
(2008).
\bibitem{haule2}K.~Haule and G.\ Kotliar  arXiv: 0805.0722.
\bibitem{dong}J. Dong {\it et al.}, 
Europhys. Lett. {\bf 83}, 27006 (2008).
\bibitem{weng}Z.-Y. Weng, arXiv: 0804.3228.
\bibitem{singh}D.\ Singh {\it et al.}, 
%and M.\ Du, 
Phys.\ Rev.\ Lett.\
{\bf 100}, 237003 (2008).
%\bibitem{felner}I.~Felner {\it et al.}, arXiv: 0805.2794.
\bibitem{lin}J.-Y.~Lin {\it et al.}, Phys.~Rev.~B {\bf 59}, 6047 (1999). 
\bibitem{petrovic}C.~Petrovic {\it et al.}, ibid.\ {\bf 66}, 054534 (2002).
\bibitem{mackenzie}A.P.~Mackenzie {\it et al.}, Phys.~Rev.~Lett.~{\bf 80}, 
161 (1998).
\bibitem{radtke}R.J.~Radtke {\it et al.}, Phys.~Rev.~B {\bf 48}, 653 (1993). 
\bibitem{hunte}F.~Hunte {\it et al.}, Nature {\bf 453},
%No.~7195, 
903 
(2008). 
\bibitem{sefat}A.~Sefat {\it et al.}, Phys.~Rev.~B {\bf 77}, 174503 (2008). 
\bibitem{chen1}G.F.~Chen {\it et al.}, 
Phys.\ Rev.\ Lett.\ {\bf 100}, 247002 (2008).
\bibitem{WHH}N.R.~Werthamer {\it et al.}, Phys.\ Rev.\ {\bf 147} 
295 (1966).


\bibitem{luetkens}H.~Luetkens {\it et al.},
Phys.\ Rev.\ Lett.\ {\bf 101}, 097009 (2008).
\bibitem{klauss}H.-H.\ Klauss {\it et al.} ibid., 077005 (2008).
\bibitem{drechsler}S.-L.~Drechsler {\it et al.}, Phys.\ Rev.\ 
Lett.\ {\bf 101} 257004 (2008).
\bibitem{graefe} H.-J.~Grafe {\it et al.}, Phys.\ Rev.\ Lett.\
{\bf 101}, 047003 (2008).

\bibitem{hess}C.~Hess {\it et al.}, erXiv:0811.1601.


\bibitem{zhu}X. Zhu {\it et al.}, Supercond.\ Sci.\ Technol.\ 
{\bf 21}, 105001 (2008).
\bibitem{remrho}Within another approach, especially to have an 
alternative determination of $B_{c2}$, where the magnetoresistance was taken 
into account, the resistance in the normal state $R_N(T)$ has been fitted 
%at a smaller $T$-range 
between $T_c$ and 
80 K resulting in $R_N(T) =7.74+1.7 \cdot 10^{-2}T^{1.4}$ given in m$\Omega$ 
and $T$ in K.
\bibitem{lab}H.~Krug {\it et al.}, Physica B {\bf 294-295}, 605  (2001).
\bibitem{orlando}T.P.~Orlando {\it et al.}, 
Phys.~Rev.~B  {\bf 19}, 4545 (1979). 
\bibitem{schossmann}M.~Schossmann and J.P.~Carbotte, ibid.~ {\bf 39}, 4210 
(1989).
\bibitem{shulga}S.V.~Shulga {\it et al.}, J.\ Low Temp.\ Phys.\ {\bf 129},
93 (2002).
 
\bibitem{jia}Y.~Jia {\it et al.} 
Appl.\ Phys.\ Lett.\ {\bf 93}, 032503 (2008).
\bibitem{weyeneth}S.~Weyeneth {\it et al.}, arXiv:0806.1024.


\bibitem{gurevich}A.~Gurevich, Physica C {\bf 456}, 160 (2007).
\bibitem{bulayevskii}L.N.~Bulayevskii {\it et al.}, Phys.~Rev.~B 
{\bf 38}, 11290 (1988).




\end{thebibliography}
\end{document}